\documentclass[12pt]{iopart}
\usepackage[symbol]{footmisc}
\usepackage{iopams}  
\usepackage{graphicx}
\usepackage{hyperref}
\usepackage{color}

\usepackage{cite}

\begin{document}

\title[]{Simulation of Deflection and Photon Emission of Ultra-Relativistic Electrons and Positrons in a Quasi-Mosaic Bent Silicon Crystal}

\author{Paulo E Iba\~nez-Almaguer$^{1}$,
Germ\'an Rojas-Lorenzo$^{1}$,
Maykel M\'arquez-Mijares$^{1}$,
Jes\'us Rubayo-Soneira$^{1}$,
Gennady B Sushko$^{2}$,
Andrei V Korol$^{2}$,
and Andrey V Solov'yov$^{2}$\footnote[1]{Corresponding author.}}

\ead{\href{mailto:paulo.ibanez@instec.cu}{paulo.ibanez@instec.cu}, % and
\href{mailto:solovyov@mbnresearch.com}{solovyov@mbnresearch.com}}

\address{$^{1}$ Instituto Superior de Tecnolog\'ias y Ciencias Aplicadas - Universidad de La Habana (InSTEC-UH), Ave. Salvador Allende 1110, 10400 La Habana, Cuba}

\address{$^{2}$ MBN Research Center, Altenh\"oferallee 3, 60438 Frankfurt am Main, Germany}

%
%\vspace{10pt}
%\begin{indented}
%\item[]August 2017
%\end{indented}

\begin{abstract}
A comprehensive numerical investigation has been conducted on the angular
distribution and spectrum of radiation emitted by 855 MeV electron and positron
beams while traversing a `quasi-mosaic' bent  silicon (111) crystal.
This interaction of charged particles with a bent crystal gives rise to various
phenomena such as channeling, dechanneling, volume reflection, and volume capture.
The crystal's geometry,  emittance of the collimated particle beams, as well as
their alignment with respect to the crystal, have been taken into account as they
are essential for an accurate quantitative description of the processes.
The simulations have been performed using a specialized relativistic molecular
dynamics module implemented in the MBN Explorer package.
The angular distribution of the particles after traversing the crystal has been
calculated for beams of different emittances as well as for different
anticlastic curvatures of the bent crystals.
For the electron beam, the angular distributions of the deflected
particles and the spectrum of radiation obtained in the simulations are
compared with the
experimental data collected at the Mainz Microtron facility.
For the positron beam such calculations have been performed for the first time.
We predict significant differences in the angular distributions and the
radiation spectra for positrons versus electrons.
\end{abstract}

\vspace{2pc}
\noindent{\it Keywords}: channeling, `quasi-mosaic' bent crystal, collimated particle beams

\submitto{\jpb}
%\maketitle

\section{Introduction}
\label{intro}

The interaction of high-energy charged particles with bent crystals
significantly depends on the relative orientation of the beam and the target.
In a crystal, the positions of constituent atoms follow certain regular pattern,
thus forming an ordered structure of the crystalline environment.
As a result, the crystal atoms can be viewed as been arranged in strings or/an
planes.
Lindhard has demonstrated \cite{Lindhard_1965} that this arrangement of atoms
produces an electrostatic field that affects charged particles passing
through a crystal at small angles with respect to crystallographic planes
(or axes)
resulting in a specific channeling motion when a particle moves along a plane
experiencing correlated interactions with the atoms.

The study of the passage of ultra-relativistic charged particle beams through
oriented crystals (including the phenomenon of channeling) has emerged as
a wide-ranging field of research \cite{Biryukov_1997,Uggerhoj_2005,ChannelingBook2014,CLS-book_2022}.
Various applications have been suggested
including beam steering \cite{Mazzolari_2014,Mazzolari_2018,Wienands_2015},
collimation \cite{Scandale_2010}, focusing \cite{Scandale_2019}, and
extraction \cite{Biryukov_1997}.
Theoretical and experimental investigations of channeling and other related
phenomena have generated valuable knowledge
\cite{Korol_2021,Haurylavets_2021,Sushko_2022,Bandiera_2013,Sytov_2016,Wistisen_2016}.
Over the last decade, a series of experiments has been conducted at various
accelerator facilities with the objective of examining channeling and radiation
emission in bent crystals \cite{Mazzolari_2014,Wienands_2015,Bandiera_2015}.

In this manuscript we present an independent analysis of
the passage of 855 MeV electrons an positrons
through an oriented bent silicon crystal and of the
radiation emitted by the particles.
The analysis is based on the results of calculations carried out within the
framework of relativistic classical molecular dynamics by means of the MBN Explorer software package
\cite{Solovyov_2012,Solovyov_2017,MBN_Explorer} and a
supplementary special multitask software toolkit MBN
Studio \cite{Sushko_2019}.
The purpose of the package is to serve as a versatile
computer program enabling the study of molecular systems
of different origin encompassing spatial scales ranging
from the atomic level to the mesoscopic domain.
The results obtained in the current simulations include
angular distributions of the particles after passing
through the crystal target and the spectral distribution
of the radiation.
In the case of electrons, our results are compared to
the experimental data presented in Ref. \cite{Mazzolari_2014} for the angular distributions
and in Ref. \cite{Bandiera_2015} for the
emission spectra.

The experiments \cite{Mazzolari_2014,Bandiera_2015} were
performed at the MAinzer MIcrotron (MAMI) facility with a highly
collimated 855 MeV electron beam.
The beam was incident on a ultra-thin silicon bent crystal.
Bending of the (111) planes was due to the ‘quasi-mosaic’ effect
\cite{Ivanov_2005,Guidi_2009,Camattari_2015} that appeared as a
result of the primary deformation create by a specially designed
holder \cite{Salvador_2018}.
After passing through the target, the electrons
were deflected by magnets and thus separated from the emitted
photons.
This enabled measurement of both the angular distribution
of the deflected electrons and the spectra of the emitted radiation.

Recently, it has been demonstrated \cite{Sushko_2022} that accounting
for specific geometry of a ‘quasi-mosaic’ bent crystal (qmBC) and
its orientation with respect to the incident beam as well as for
the beam's transverse size and emittance is essential for
an accurate quantitative description of experimental results on the
beam deflection by such crystals.
In the current simulations, these parameters have been accounted for
along with the crystal thickness and curvature radii.
The simulations have been performed for both electron and positron
beams with similar characteristics.
The predictions made for the positrons are highly relevant
to the experiments with the positron test beam which is planned
to be constructed at MAMI within the framework of the ongoing
European project TECHNO-CLS \cite{TECHNO-CLS}.
This project aims at the breakthrough in technologies
require for the practical realisation of novel
crystal-based gamma-ray Light Sources
that can be constructed through exposure of oriented crystals to the
beams of ultra-relativistic charged particles
\cite{CLS-book_2022,KorolSolovyov:EPJD_CLS_2020}.

This contribution is organized as as follows: In the next section, we outline the methodology utilized in the simulations as well as the parameters of the crystal target and its alignment with respect to the beam following the description of the experimental setup \cite{Mazzolari_2014,Bandiera_2015}. The outcomes of the calculations are analyzed in Results and Discussion section, and a summary of the results obtained is given in Conclusions.

%%%%%%%%%%%%%%%%%%%%%%%%%%%%%%%%%
\section{Methodology \label{Theory}}

In this paper we employ the method of relativistic classical molecular dynamics \cite{Sushko_2013}, which is implemented in the MBN Explorer package, to simulate the motion of ultra-relativistic charged particles
within the electrostatic field of a crystalline medium. This approach involves generating a significant number $N$ of statistically independent trajectories for projectile particles. These trajectories can then be further analyzed to provide a quantitative characterization of the particles' motion and the radiation they emit.

To model the motion of an ultra-relativistic particle with mass $m$, charge $q$, and energy $\epsilon$ in a medium, the following relativistic equations of motion are integrated:
\begin{equation}
	\left\lbrace
	\begin{array}{l}
		\dot{\vec{r}} = \vec{v}\\
		\dot{\vec{p}} =	q\vec{E} 
	\end{array}
	\right. .
	\label{eq:01} %{Equations:eq.01}
\end{equation}

The instantaneous position and velocity of the particle are
${\vec{r}} = \vec{r}(t)$ and $\vec{v}= \vec{v}(t)$ respectively.
A dot above the letter denotes differentiation with respect to
time $t$.
The momentum $\vec{p}$ in terms of velocity reads
$\vec{p}=m\gamma \vec{v}(t)$ where $\gamma =\epsilon /mc^2$ stands
for the relativistic factor, where $c$ is the speed of light in
vacuum.

The electric field in the point $\vec{r}$ is calculated as
$\vec{E}(\vec{r})=-\nabla _{\vec{r}} \phi(\vec{r})$ with
$\phi (\vec{r})$ standing for the field's potential.
This quantity is calculated as the sum of potentials of individual
atoms, which contribution is not negligible.
The atomic potentials can be computed within the frameworks of the
approximations due to Moli\`{e}re \cite{Moliere_1948}
and Pacios \cite{Pacios_1993}.

To ensure that all trajectories are statistically independent and
each one corresponds to a unique crystalline environment, the
positions of the atoms are generated accounting for random
displacement from the nodes due to thermal vibrations.
In addition to this, initial transverse coordinates and velocities
are generated randomly, with their distribution determined by the
transverse size and divergence of the beam.

Silicon crystal has the diamond cubic structure and thus its
mechanical properties are highly anisotropic.
As a result, when the crystalline medium is subjected to two moments
of force to achieve primary curvature, some secondary curvatures may
arise within the solid.
A well-known secondary deformation results in
the anticlastic curvature
that occurs in the perpendicular direction with respect to the
primary curvature.
When the two curvatures are combined, the deformed crystal acquires
the shape of a saddle.
Another type of the deformation caused by anisotropic effects is the
‘quasi-mosaic’ (QM) curvature \cite{Ivanov_2005}.
The qmBCs belong to a class of bent crystals featuring two curvatures
of two orthogonal crystallographic planes.
In the silicon crystals used in the experiment
\cite{Mazzolari_2014,Bandiera_2015} the (111) planes were bent
due to the QM effect.

\begin{figure}
\centering 
\includegraphics[width = 9.5cm]{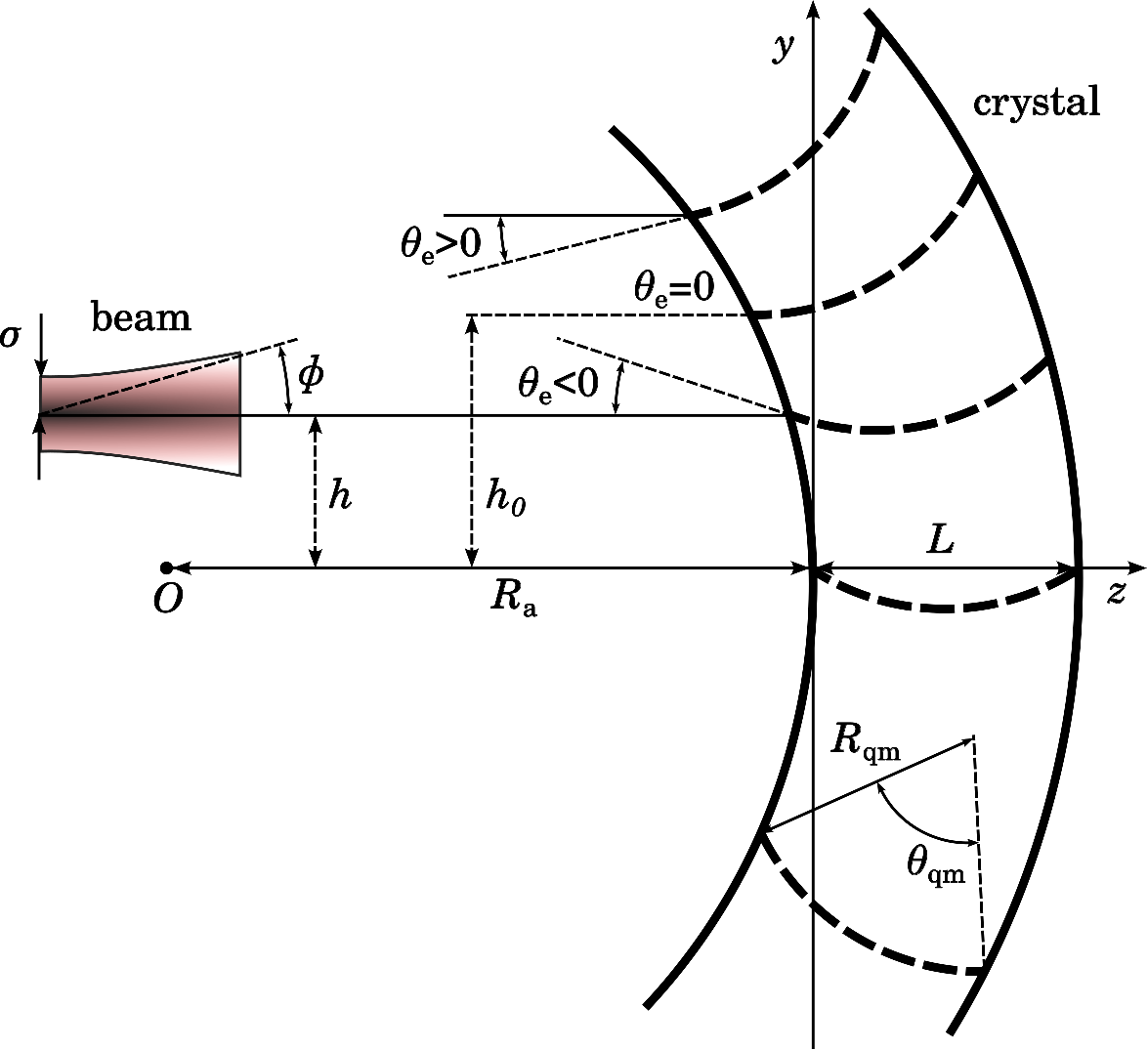}
\caption{Geometry of a ‘quasi-mosaic’ bent crystal plate of a
thickness $L$ and its alignment with respect to an incident beam
of a transverse size $\sigma$ and angular divergence $\phi$.
The point $O$ denotes the center of anticlastic curvature of radius
$R_{\mathrm{a}}$.
The Si(111) planes (thick dashed curves) are bent due to the QM
effect with the curvature radius
$R_{\mathrm{qm}}$; $\theta _{\mathrm{qm}}=L/R_{\mathrm{qm}}$
stands for the QM bending angle.
Other notations are explained in the text.}
\label{fig:1}
\end{figure}

Following Ref. \cite{Sushko_2022}, Fig. \ref{fig:1} illustrates
the alignment of a qmBC and an incident beam.
It is important to note that the representation is merely a
schematic and not to scale.
The anticlastic radius $R_\mathrm{a}$  is on the order of meters,
and the QM radius $R_\mathrm{qm}$ is on the order of centimeters,
while the thickness $L$ is on the order of tens of micrometers.
Therefore, the actual curvatures are not as pronounced as
they are shown in the figure.

%%%%%%%%%%%%%%%%%
%%%%%%%%%%%%%%%%%%

In the coordinate system chosen, the $z$-axis is aligned with
the incident beam direction which is normal to the face of the not
deformed (i.e., not bent) crystal plate.
The primary curvature (not shown in the figure) causes the plate
to bend around the $y$ direction towards the positive direction of
the $z$-axis.
The anticlastic curvature of radius $R_{\mathrm{a}}$ bends the
plate around the $x$ direction towards the negative $z$-axis.
In Fig. \ref{fig:1} the point $O$ denotes the center of anticlastic
curvature.
The QM bending of the (111) crystallographic planes is around
the $x$ direction ($R_{\mathrm{qm}}$ denotes the QM bending radius).
In the present paper the case of a planar channeling is considered.
Therefore, in the simulations, the $z$-axis direction has been
chosen well away from major crystallographic axes to avoid axial
channeling.

The entrance angle $\theta _{\mathrm{e}}$ between the incident beam
direction and the nearest QM bent plane at the entry of the crystal
depends on the $y$-coordinate $h$ of the center of the beam,
$\theta _{\mathrm{e}}(h) = (h-h_0)/R_{\mathrm{qm}}$.
The quantity
$h_0=\theta _{\mathrm{qm}}R_{\mathrm{a}}/2$ stands for the
displacement at which the entrance angle is zero.
A probability of a particle to be accepted into the channeling mode at the crystal entrance becomes significant if
$\theta_{\rm e}$ does not exceed Lindhard's critical angle
$\theta_{\rm L}$.
For an ideally collimated beam ($\phi=0$) this condition
is met for the particles that enter the crystal within the
the interval $h_0 \pm \theta_{\rm L} R_{\rm a}$.
The projectiles that are accepted at $y=h$ and channel
through the whole crystal are deflected by the angle
lying within the interval $\theta_{\rm e}(h) + \theta_{\rm qm}
\pm \theta_{\rm L}$ \cite{Sushko_2022}.

The particles entering the crystal in the region $-h_0 < h < h_0$ (i.e.
with $\theta _{\mathrm{e}}>0$) can experience either volume
capture \cite{Taratin_1986} or volume reflection \cite{Taratin_1987}
of the curved crystalline planes during their propagation through
the crystal volume.
These events take place at the points at which
a particle's trajectory becomes tangent to the planes.
For a given value of $h$ the point of the volume
capture and the volume reflection is positioned at the
distance $L/2 - R_{\rm qm} h/R_{\mathrm{a}}$ from the entrance point
to the crystal.
This distance is equal to zero if $h = h_0$ and
to $L$ for $h=-h_0$.
Particles that move in the channeling regime after the
volume capture  exit the crystal at the angle
$\theta_{\rm s}
=\theta_{\rm e}(h)+\theta_{\rm qm}/2+h/R_{\mathrm{a}}$.
In the event of volume reflection particles are deflected by some
characteristic angle $\theta_{\rm vr}$ which does not depend on
the choice of $h$ but is determined by radius $R_{\rm qm}$
and the particle energy.
After the volume reflection particles experience multiple scattering
within the remaining crystal volume and exit the crystal
at the characteristic angle $\theta_{\rm vr}$.

If at the entrance $h > h_0 + \theta_{\rm L} R_{\rm a}$ then
the particle is neither accepted nor volume reflected but
experiences multiple scattering which becomes closer to the
scattering in the amorphous medium as $h$ increases.

To reproduce the geometry shown in Fig. \ref{fig:1} with the MBN
Exlporer and MBN Studio packages,
the first step is to generate a non-deformed crystalline medium
within the spatial region from $z=0$ to $z=L$ \cite{Sushko_2013}.
Then, two transformations can be applied to the silicon crystal
structure:
(i) ‘quasi-mosaic’ bending of the Si(111) planes,
and (ii) anticlastic bending of the planes parallel to the
$xy$-plane \cite{Sushko_2022}.
Since the crystal has a finite thickness, the equations of
motion Eq. \ref{eq:01} can be integrated from some point
$z=-\Delta z$ before the entrance to the crystal up to
$z=L+\Delta z$ which is beyond the exit point $z=L$.
Doing this, one accounts for the particle--crystal-atoms
interaction as it approaches the crystal, passes though it
and moves away from the crystal.
Typically, the value of $\Delta z$ equals to the
cutoff radius $\rho_{\max}$ for the potential chosen
to describe the particle--atom interaction.
This potential decreases rapidly with increasing distance
from the atom.
Therefore, when calculating the potential acting on the particle in
the crystal or in its vicinity, one can account only for those atoms
that are located inside a sphere of the radius $\rho_{\max}$
with the center at the instant position of the particle.

The simulation parameters have been selected according to the
experiment \cite{Bandiera_2015}:
The crystal thickness is $L = 30.5$ $\mu$m and
the QM bending radius $R_{\mathrm{qm}}=3.35$ cm.
The anticlastic radius was not specified in the cited paper.
Basing on the value $R_{\mathrm{a}}=3.66$ m obtained in the
experimental measurements \cite{Guidi_2009} of the quasi-mosaic
crystals bent with smaller QM radii
($R_{\mathrm{qm}}\approx 1.5-1.8$ cm),
in the current simulations two anticlastic radii
$R_{\mathrm{a}}=$ 5 and 10 m have been considered.

The simulations have been carried out for 855 MeV Gaussian beams
of the transverse size $\sigma = 50$ $\mu$m and of two
divergences $\phi=10$ and $20$ $\mu$rad.
% Two values of emittance ($\sigma \times \phi$) were selected to
% analyze the effects of the divergence on the angular distribution.
The passage of the particles has been simulated for the following two
different beam–crystal geometries
(see Fig.  \ref{fig:1}):
(i) $y=h_0$ which corresponds to beam alignment with
bent Si(111) planes at the entrance ($\mathrm{\theta_e}=0$),
(ii) $y=h < h_0$ with $h$ corresponding to the
entrance angle $\theta_{\rm e}=-10$ $\mu$rad.

\begin{figure}
\centering
\includegraphics[width = 9.5cm]{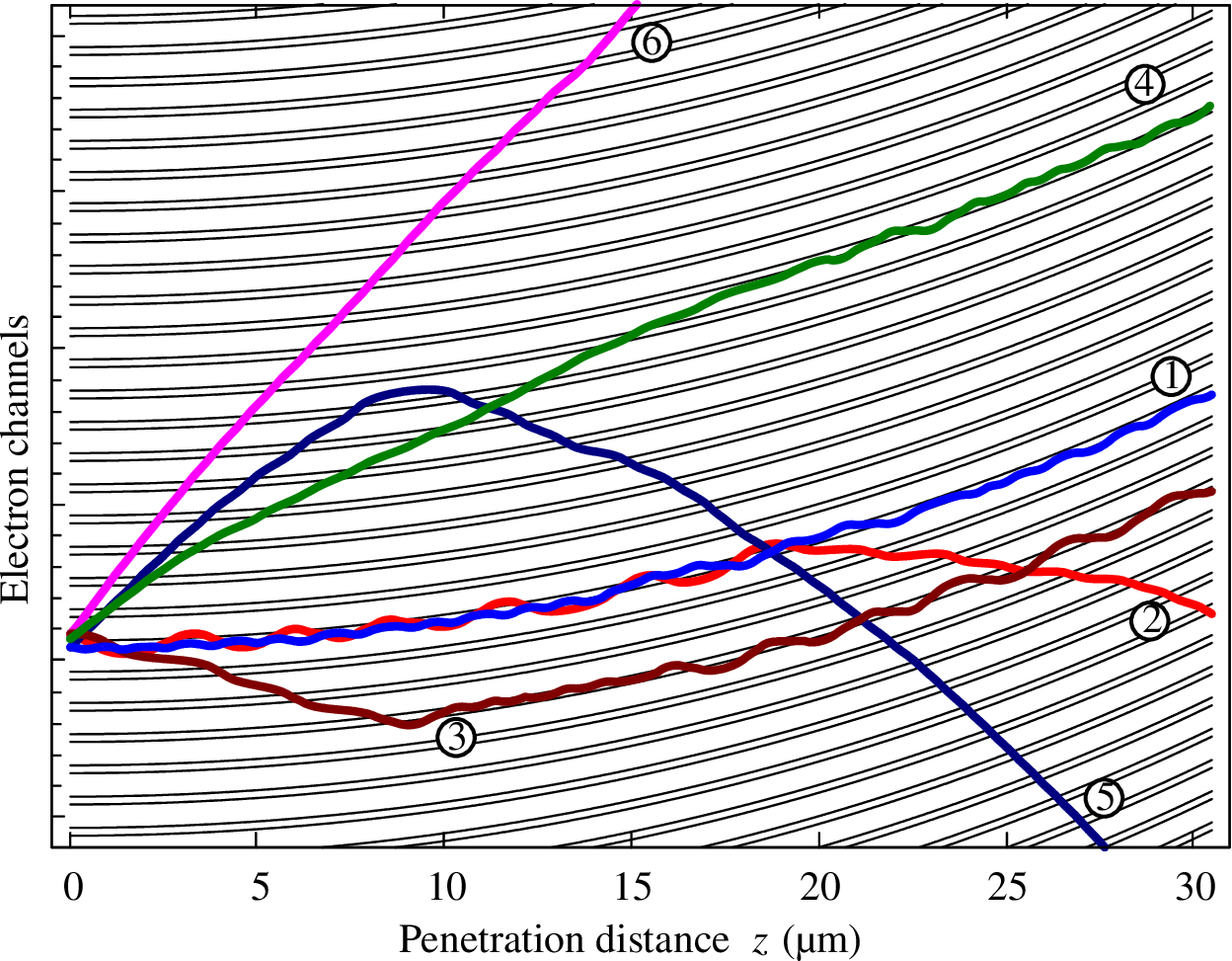}
\caption{Selected trajectories showcasing processes associated with the interaction of electrons with a bent crystal.
The thin lines represent the bent Si(111) planes
and the thick lines represent the projection in the
$yz$-plane of the simulated electron trajectories.}
\label{fig:2}
\end{figure}

Sets of statistically independent trajectories have been
calculated once the geometry an the beam characteristics were
defined.
Figure \ref{fig:2} displays several simulated trajectories
selected to represent processes related to the interaction of an
electron beam with the crystal.
Thin lines show the $(y,z)$ cross sections of the bent Si(111)
planes separated with alternate wide ($d_{\rm W}=2.352$ \AA) and
short ($d_{\rm S}=0.784$ \AA) spacings.
The electron channels of the width
$d=d_{\rm W}+d_{\rm S}=3.136$ \AA{} are
defined as the regions between the centerlines of the two nearest
wide spacings.
Tick marks shown on the left vertical axis of the figure
mark the channels' boundaries at the crystal entrance.

%%%%%%%%%%%%%%%%%
%%%%%%%%%%%%%%%%

When the entrance angle is smaller than Lindhard's critical angle
$\Theta_{\rm L}$
(maximum incident angle consistent with the channeling condition
\cite{Lindhard_1965}) particles may be accepted into a channel.
Within the continuous potential approximation one writes
$\Theta_{\rm L} = (2U_0/\epsilon)^{1/2}$ with $U_0$
standing for the depth of the continuous interplanar potential.
Using the value $U_0 \approx 24$ eV for the Si(111)
channel (see, e.g., Ref. \cite{HaurylavetsEtAl:NIMA_v1058_168917_2024})
one calculates $\Theta_{\rm L} \approx 0.24$ mrad for
a 855 MeV projectile.
An accepted particle can traverse the entire crystal in the
channeling mode (channeling: trajectory 1),
or it may exit the channel at some point in the crystal bulk
(dechanneling: trajectory 2).
A non-accepted particle starts its motion in the over-barrier
mode but can be captured into the channelling mode afterwards (rechanneling: trajectory 3).
 
When the incident angle of a particle is greater than
$\Theta_{\rm L}$ it starts to move in the over-barrier mode.
Due to the crystal bending, at some point inside the crystal
particle's velocity may become tangent to the crystal plane
resulting in the particle's capture into the channeling mode
(volume capture: trajectory 4).
Another scenario for an over-barrier particle, which moves under a
small angle to the bent crystal, implies a reflection to
the side opposite the bend (volume reflection \cite{Taratin_1987}:
trajectory 5).
Finally, a non-accepted particle can pass through
the whole crystal in the over-barrier mode (trajectory 6).
 
The simulated trajectories, being statistically independent,
can be used to calculate spectral and angular distributions of
the emitted radiation.
The averaged spectral distribution of energy emitted within
the cone $\theta \leq \theta _0$ with respect to the incident beam
is computed as follows:
\begin{eqnarray}
\left\langle
\frac{dE(\theta \leq \theta _0)}{d\omega}\right\rangle
=
\frac{1}{N} \sum ^N_{n=1}\int _0^{2\pi}d\phi \int _0^{\theta _0}\theta d\theta \frac{d^3 E_n}{d\omega d\Omega }\ .
\label{eq:02} %{Equations:eq.01}
\end{eqnarray}
Here, $\omega $ stands for the frequency of radiation,
$\Omega$ is the solid angle corresponding to the emission
polar angles $\theta $ and $\phi$,
and ${d^3 E_n}/{d\omega d\Omega} $ is the spectral-angular
distribution of radiation emitted by a projectile
moving along the $n$th trajectory,
$N$ stands for the total number of simulated trajectories.
The numerical procedures implemented in MBN Explorer package to
calculate these distributions \cite{Sushko_2013} are based on the
quasi-classical formalism due to Baier and Katkov \cite{Baier_1998}.
This method combines classical description of the motion in an
external field with the quantum corrections due to the radiative
recoil.

The resulting spectrum accounts for channeling radiation owing to the
motion of particles in the channeling mode, coherent and incoherent
bremsstrahlung due to over-barrier motion, and
the synchrotron-type
radiation as result of the circular motion along an arc in a bent
crystal.

%%%%%%%%%%%%%%%%%%%%%%%%%%%%%%%%%
\section{Results and Discussion \label{Results and Discussion}}

%%%%%%%%%%%%%
\subsection{Angular distribution of electrons}

%%%%%%%%%%%%%%%%%
\begin{figure}
\centering 
\includegraphics[width = 9.5cm]{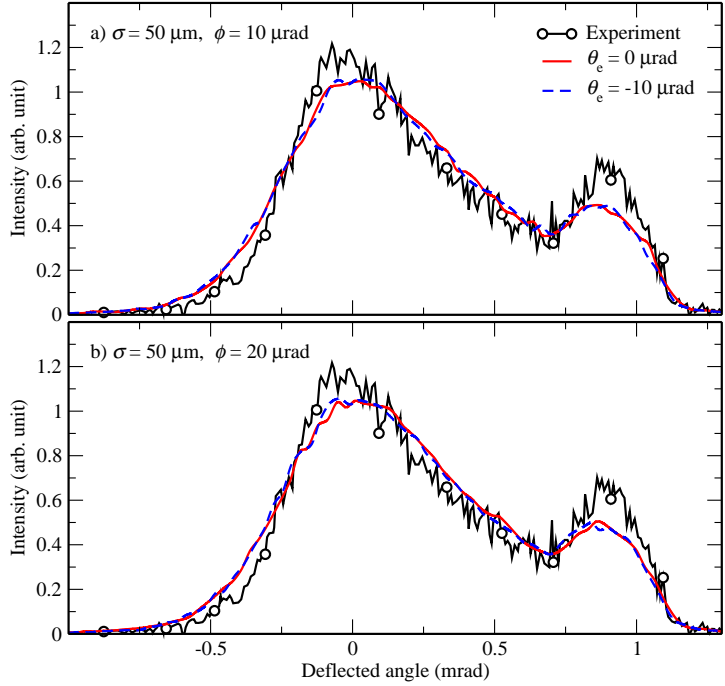}
\caption{
Angular distributions of deflected electrons.
The simulations (lines without symbols)
refer to a 855 MeV beam of the transverse size
$\sigma=50$ $\mu$m
and angular divergences of $\phi=10$ $\mu$rad (panel a)
and $\phi=20$ $\mu$rad (panel b)
incident on a quasi-mosaic Si(111) crystal
with anticlastic radius $R_\mathrm{a}=5$ m.
Solid lines correspond to the distribution calculated for  the
 entrance angle
$\theta_\mathrm{e}=0$, dashed lines correspond to
$\theta_\mathrm{e}= -10$ $\mu$rad.
The lines with circles stand for the experimental data reported in Ref. \cite{Mazzolari_2014}.
All dependencies are normalized to the unit area.
}
\label{fig:3}
\end{figure}
%%%%%%%%%%%%%%%

The trajectories have been simulated utilizing atomic potentials
derived from the Moli\`{e}re and the Pacios models for the
electron--atom interaction.
The results obtained for the angular distribution of the deflected
electrons after interaction with the crystal turned out to be similar
for both potentials.
For the spectral distribution of the radiation it has been found
that the calculations with the Pacios potential exhibit better agreement with the experimental results.
This feature is in line with the analysis carried out earlier
in Ref. \cite{Haurylavets_2021}.

The angular distributions calculated for the bent crystal with
the anticlastic radius  $R_\mathrm{a}=5$ m are shown in
Fig, \ref{fig:3}.
The simulate data refer to two entrance angles
(see Fig. \ref{fig:1}):
$\theta_\mathrm{e} = 0$ (solid lines without symbols)
and $\theta_\mathrm{e} =-10$ mrad (dashed lines).
Two panels  present the distributions simulated for different
values of the beam divergence: $\phi=10$ $\mu$rad (panel a)
and $\phi=20$  $\mu$rad (panel b).
In both cases the beam size is fixed at $\sigma=50$ $\mu$m.
The experimental data shown were obtained by digitizing
the graphical data presented for the channeling alignment
in Figure 3 from  Ref. \cite{Mazzolari_2014}.

The simulated and measured angular distributions have the
characteristic pattern of two well pronounced peaks
interlinked by an intermediate region.
The peak on left side is primarily contributed by the electrons
moving in the over-barrier mode from the entrance point up to the
crystal exit.
These particles experience multiple scattering from the crystal
atoms which leas to the broadening of the angular distribution
at the exit as compare to the angular divergence of the incident
beam.

The main contribution to the right-side peak, which is centered
at the QM bending angle $\theta_{\rm qm}=L/R_{\rm qm}=0.91$ mrad
(see Fig. \ref{fig:1}), is due to the electrons that channel
through the entire crystal length.
Additionally, electrons that are trapped in the
channeling mode somewhere inside the crystal
contribute to the distribution in the vicinity of this peak.
The intermediate region is predominantly formed by electrons
that are accepted in the channeling mode at the entrance but
then dechannel at some point in the bulk.

Figure \ref{fig:3} shows that there are no significant
differences between the distributions obtained for the
the two entrance angles $\theta_{\mathrm e}$.
In both cases,  the peak values are smaller than the experimentally
measure ones.
This is because the simulation show higher contribution from the
dechanneling process compared to the experiment.
Consequently, more electrons are deflected in the intermediate
region resulting in the decrease in peak values and
causing the peaks shift towards each other.

%%%%%%%%%%%%%%%%%%%%%%%%%%%%%%%
\begin{figure}
\centering 
\includegraphics[width = 9.5cm]{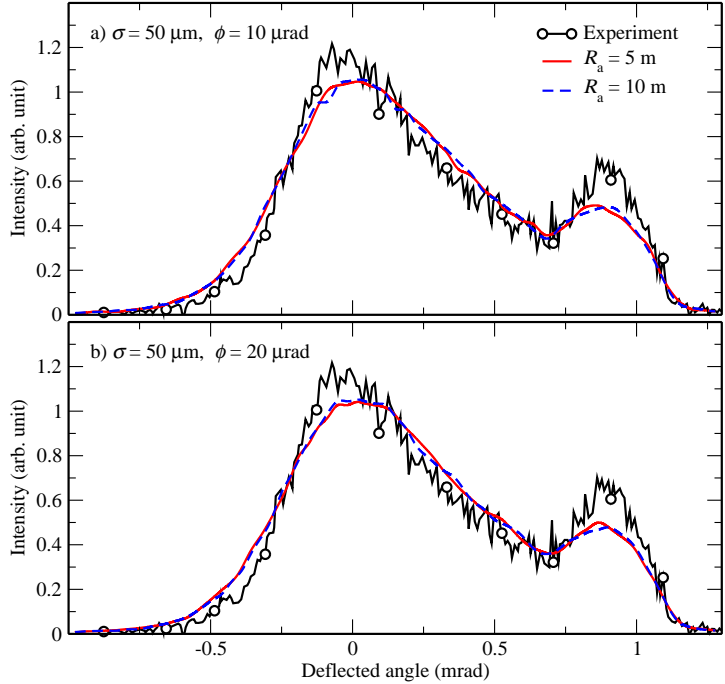}
\caption{
Angular distribution of the deflected electrons simulated
for different values of the beam divergence
($\phi=10$ and 20 $\mu$rad in the top and bottom graphs,
correspondingly) and for two values of the anticlastic curvature
radii  $R_\mathrm{a}$ as indicated in the common legend shown in the
top graph.
All dependencies shown correspond to the beam transverse size
$\sigma=50$ $\mu$m an to the incident angle
$\theta_{\mathrm e}=0$.
Solid line with open circles corresponds to the experimental
data reported in Ref. \cite{Mazzolari_2014}.
All distributions are normalized to the unit area.
}
\label{fig:4}
\end{figure}
%%%%%%%%%%%%%%%%%%%%%%%%%%%%%%%

The angular distributions of deflected electrons simulated for two
values of the anticlastic curvature radius,
$R_{\mathrm a}=5$ and 10 m, are compared in Fig. \ref{fig:4}.
The simulations refer to the beam of the
size $\sigma=50$ $\mu$m incident on the crystal at
$\theta_{\mathrm e}=0$.
The top graph corresponds to the beam divergence $\phi=10$ $\mu$rad,
the bottom graph -- to $\phi=20$ $\mu$rad.
The experimental data taken from Ref. \cite{Mazzolari_2014}
are also shown.

The angular distributions obtained for the two values of
anticlastic curvature radii are similar,
suggesting that for large values of $R_{\mathrm a}$,
the fraction of deflected electrons at each angle
is independent of this parameter.
Comparing the distributions shown in the top and bottom graphs,
one concludes that they are weakly dependent on the beam
divergence.
This is understandable since the divergences considered
are much smaller than Lindhard's critical angle
$\theta _\mathrm{L} \approx 240$ $\mu$rad.

%%%%%%%%%%%%%%%%%%%
\subsection{Angular distribution of positrons}

\begin{figure}
\centering 
\includegraphics[width = 9.5cm]{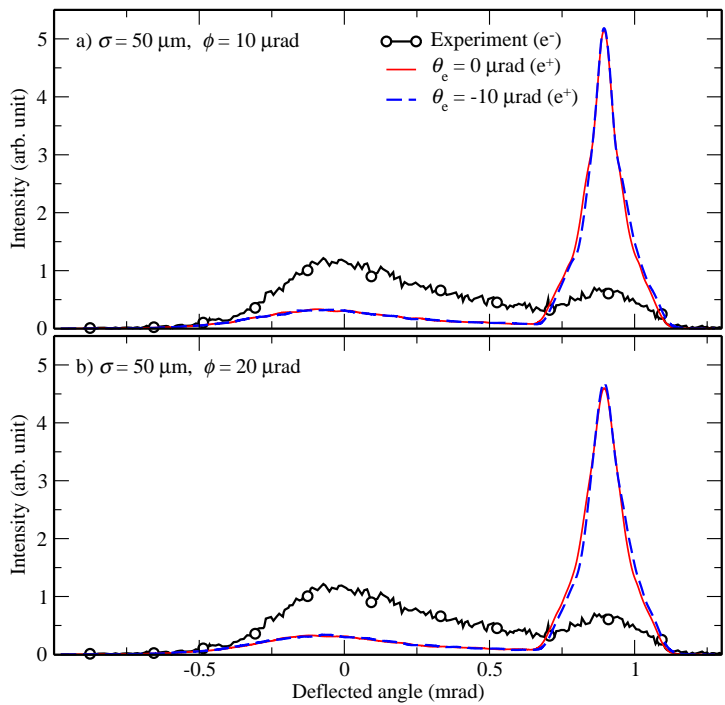}
\caption{
Similar as in Fig. \ref{fig:3} but for a 855 MeV
positron beam incident on the quasi-mosaic Si(111)
crystal.
All simulated distributions refer to the anticlastic
curvature radius  $R_\mathrm{a}=5$ m.
The experimental data for electrons
\cite{Mazzolari_2014} are also shown.
}
\label{fig:5}
\end{figure} 

Angular distributions of positrons deflected by the
same quasi-mosaic bent silicon crystal has also been analyzed.
Figure \ref{fig:5} shows the distributions obtained in the
simulations of the 855 MV positron beams
with transverse size $\sigma = 50$ $\mu$m and
divergences $\phi=10$ and $20$ $\mu$rad (top and bottom graphs,
respectively) incident on the bent Si(111) crystal
with the anticlastic curvature radius $R_\mathrm{a}=5$ m.
As in the case of electrons (see Fig.  \ref{fig:3}) two incident
geometries have been considered corresponding to the
the entrance angles
$\theta_\mathrm{e} = 0$ and $-10$ mrad.
The experimentally measured distribution of electrons
\cite{Mazzolari_2014} is also shown for the sake of comparison.
 
Most of the positrons pass through the whole crystal
moving in the channeling mode.
As a result, the channeling peak at
$\theta_{\rm qm}=0.91$ mrad
is much more pronounced than in the electron distribution.

The positrons that are not accepted in the channeling regime
at the entrance (as well as those that are accepted but dechannel
shortly after) and propagate throughout the entire crystal
in the over-barrier mode exhibit a behavior similar to that
of the electrons.
For positrons, the left peak is centered at a larger negative
angle than the peak in the electron distribution.
This indicates that relative fraction of volume reflected
positrons (which contribute to the distribution in the domain of
negative deflection angles) is larger than that for electrons.
The positron dechanneling rate is much less than the
dechanneling rate of electrons of the same energy.
Therefore, the channeling peak in the positron distribution is
enhanced greatly whereas the left-peak value is notably smaller that
in the electron distribution.

As in the case of electrons, the angular distributions of
deflected positrons are similar for the two investigated entrance angles, see Fig. \ref{fig:5}.
However, for positrons, the impact of the beam divergence on the distribution is more pronounced.
This is clearly seen if comparing the channeling peak values
in the top an bottom graphs:
for the beam with smaller divergence (top) the channeling peak
is ca 10 \% higher than for the wider beam (bottom).

\begin{figure}
\centering 
\includegraphics[width = 9.5cm]{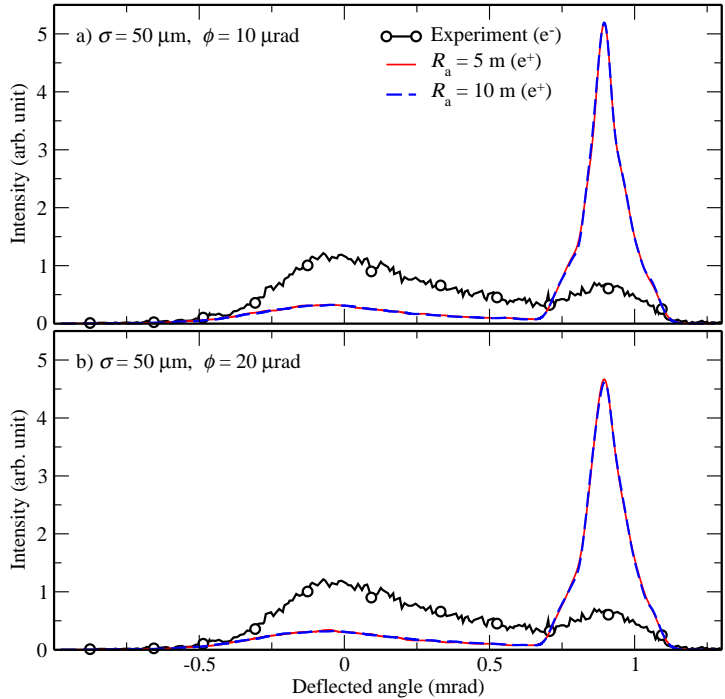}
\caption{
Similar as in Fig. \ref{fig:4} but for 855 MeV
positron beams incident on the quasi-mosaic Si(111)
crystal.
The experimental data for electrons
\cite{Mazzolari_2014} are also shown.
}
\label{fig:6}
\end{figure}

Angular distributions simulated for the positron beams
centered at a $y=h_0$
(corresponding to tangential geometry at the crystal entrance,
see Fig. \ref{fig:1})
incident on Si(111) crystals with anticlastic radii
$R_{\mathrm{a}}=5$ and 10 m are shown in Fig. \ref{fig:6}.
Two values of the beam emittance have been considered
while maintaining constant value of the beam's transverse size.
The distribution measured in the experiment with electrons
\cite{Mazzolari_2014} is shown for comparison.

%%%%%%%%%%%%%%%%%%
%%%%%%%%%%%%%%%%%%%

The results from this figure corroborate the fact that for large
anticlastic radii, the angular distribution of charged
particles becomes independent on it.
Furthermore, for both radii, the channeling peak
becomes less intensive and broader as the
value of the beam divergence increases.

The atomistic approach implemented in MBN Explorer allows one
to look for and to identify various processes mentioned
earlier (see  Fig. \ref{fig:2}) for each simulated trajectory.
This enables a more detailed analysis of the contribution
of these processes to the overall distribution.

% Positrons and electrons are deflected in the same range of angles,
% but each process has a different weight in the overall distribution
% due to their distinct dechanneling lengths \cite{Kumakhov_1989}.
% The majority of positrons channel through the entire crystal,
% while electrons are dechanneled at some point within the crystal.

%%%%%%%%%%%%%%%%%%%%%%%%%%%%%%%%%%%%%%%%
\subsection{Spectra of emitted radiation}

\begin{figure} 
\centering 
\includegraphics[width = 9.5cm]{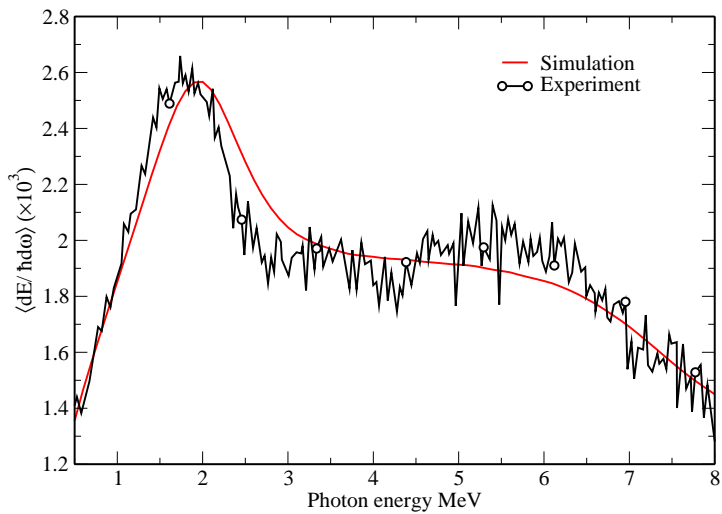}
\caption{
Spectral distribution of radiation emitted by
855 MeV electrons passing through the
‘quasi-mosaic’ bent Si(111) crystal.
Solid (red) curve without symbols stands for the results of the
current simulations; solid (black) curve with symbols corresponds
to the experimental data reported in Ref. \cite{Bandiera_2015}.
Note that the values of
$\left\langle dE/d(\hbar\omega)\right\rangle$
are shown being multiplied by the factor $10^3$.
The intensity of the of background radiation due to the
incoherent bremsstrahlung is approximately $0.35\times 10^{-3}$.
}
\label{fig:7}
\end{figure}

MBN Explorer allows one to calculate spectral and angular
distributions of the radiation emitted by beam particles
using the scheme presented by  Eq. (\ref{eq:02}).
For each simulated trajectory
(the index $n=1,2,\dots, N$ enumerates the trajectories)
the spectral-angular distributions
$d^3E/d\omega\,d\Omega$
are calculated for specified values of the
emitted photon energy $\hbar\omega$ and emission angle $\theta$
(by default, the full range from 0 to $2\pi$ is assumed
for the polar angle $\varphi$).
Integrating numerically the individual spectral-angular
distributions over $\theta$ within a specified cone $\theta_0$
with respect to the incident beam and carrying out averaging
over all trajectories one calculates averaged spectral
distribution $\left\langle dE/d(\hbar\omega)\right\rangle$
of radiation emitted by the beam.

In this section, the distributions
$\left\langle dE/d(\hbar\omega)\right\rangle$ presented
have been calculated for 855 MeV electron and positron beams
incident on the bent Si crystal being aligned with the (111) at
the entrance (i.e., $\theta_\mathrm{e}=0$).
The data refer to the anticlastic curvature radius
$R_{\mathrm{a}}=5$ m, and to the beam
transverse size $\sigma=50$ $\mu$m and divergence
$\phi=20$ $\mu$rad.
These values of $\sigma$ and $\phi$ correspond to the
emittance $\sigma\times\phi$ of the electron beam used
in the experiment \cite{Bandiera_2015} where the spectral
distribution of the emitted radiation was measured.
The calculated spectra presented below correspond to the
the emission cone $\theta_0$ equal to
$\theta_{\rm qm}+3/\gamma \approx 2.7$ mrad.
This cone is wide enough to collect almost all emission radiated
from each simulated trajectory.
Its value is slightly larger than the cone $\theta_0\approx 2.4$ mrad
use in the experiment.

Figure \ref{fig:7} presents the calculated spectral distribution
of radiation (red solid line without symbols) emitted by
the electron beam and the experimentally measured spectrum
(black solid line with open circles) \cite{Bandiera_2015}.
The experimental values has been obtained by digitalizing
the "CR-exp" curve in figure 2 in the cited paper.
The experimental data were reported in arbitrary units,
whereas the current simulations produce absolute values.
Therefore, to carry out the comparison both results the experimental
data has been re-scaled to produce the same area as in the
simulations.

By comparing the curves in Fig. \ref{fig:7}
one can conclude that there is a good agreement between the
experimental and
theoretical results both in terms of the spectral distribution shape
and absolute values.
In the photon energy interval presented, $\hbar\omega=0.5 -8$ MeV,
the spectrum is dominated by the radiation produced by the
channeling electrons, i.e. by the channeling radiation (ChR).
The peak value of ChR at about $1.8$ MeV is approximately 7 times
larger than the intensity of the incoherent bremsstrahlung
background (not shown in the figure).

\begin{figure}
\centering 
\includegraphics[width = 9.5cm]{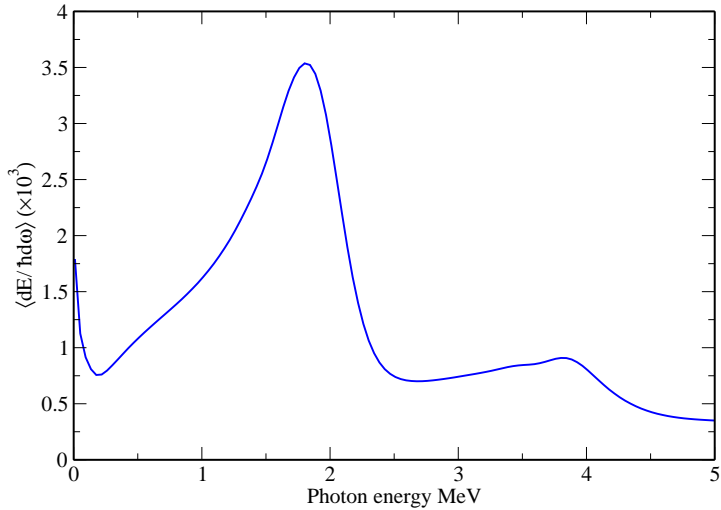}
\caption{
Spectral distribution of radiation emitted by
855 MeV positrons passing through the ‘quasi-mosaic’ bent Si(111)
crystal.
Note that the values of
$\left\langle dE/d(\hbar\omega)\right\rangle$
are shown being multiplied by the factor $10^3$.
}
\label{fig:8}
\end{figure}

The spectrum of radiation emitted by the positron is shown
in Fig. \ref{fig:8}.
In the photon energy interval $\hbar\omega \approx 1-2$ MeV
the spectrum is dominated by the peak of ChR, which is more
powerful and narrower than in the electron case.
The profile of the peak
is similar to the profile of the emission into the
fundamental harmonic in a planar undulator,
in which an ultra-relativistic charge moves
along harmonic trajectory (see, e.g., \cite{ChannelingBook2014} and
references therein).
It is not surprising, since the channeling oscillations of positrons
are nearly harmonic (see, e.g., examples of simulated positron
trajectories presented in Refs. \cite{SushkoEtAl:JPConfSer_v438_012018_2013,Korol_2021}).
The second peak at $\hbar\omega \lesssim 4$ MeV, although much less
intensive and much broader, corresponds to the emission into the second
harmonic of ChR.
Channeling oscillations of electrons are strongly
anharmonic.
As a result, their ChR is spread over broader interval of the photon
energies making the spectral distribution less similar to that
intrinsic to the undulator radiation.

%%%%%%%%%%%%%%
%%%%%%%%%%%%%%%%%%%%%%%%%

The feature that is clearly seen in the positron
emission spectrum is due to a circular motion of the channeling
particles.
This motion leads to the emission of a synchrotron-type radiation (SR),
which contributes to the low-energy part of the spectrum,
$\hbar\omega \lesssim  0.2$ MeV.
This feature, predicted theoretically in Ref. \cite{KaplinVorobiev:PLA_v67_p135_1978},
has been recently analysed quantitatively by means of
relativistic molecular dynamics simulations
\cite{SushkoEtAl:JPConfSer_v438_012018_2013,
HaurylavetsEtAl:NIMA_v1058_168917_2024,ShenEtAl:NIMB_v424_p26_2018}.
The characteristic energy $\hbar\omega_{\rm c}$ (in MeV) beyond which
intensity of SR decreases rapidly
\cite{Jackson} can be written as follows:
$\hbar\omega_{\rm c} \approx 0.22\epsilon^3 /R$ with $\epsilon$ in GeV
and the curvature radius $R$ in cm.
For a 855 MeV projectile and for $R=R_{\rm qm} = 3.35$ cm this
estimate produces $\hbar\omega_{\rm c}  \approx 50$ keV.

We note that the mechanism of the SR emission by a channeling particle
does not depend on the sign of its charge.
However, as a rule the dechanneling length for a positron is much larger
than for an electron of the same energy, therefore, the SR emitted by
a channeling positron is more intensive as it can experience circular
motions over larger distances channels.
In the electron emission spectrum, Fig.  \ref{fig:7}, the increase
in the low-energy part of the spectrum is not seen because its stars
below the lowest photon energy of 0.5 MeV shown in the figure.
This cut-off has been introduced to match the interval of photon
energies investigate in the experiment \cite{Bandiera_2015}.

%%%%%%%%%%%%%%%%%%%%%%%%%%%%%%%%%
\section{Conclusion \label{Conclusion}}

The atomistic approach of the MBN Explorer allows one to monitor
the changes in the distribution of electrons and positrons
deflected by a 30.5 $\mu$m thick quasi-mosaic bent Silicon
crystal upon the parameters of the beam and the crystal bending,
as well as to obtain the spectrum of emitted radiation.

The simulation results for the 855 MeV electron beams of different
emittances and orientations with respect to the crystal have been
compared with experimental data.
A good agreement of the simulations withe the experiment can be stated.
Some discrepancies have been noted in the angular distribution
of deflected electrons.
Possible reasons for the discrepancies can be attributed to
the particular force fields (Moli\`{e}re and Pacios )
chosen to describe the particle-atom interaction in the course of
simulations, and to the phenomena not accounted for
(e.g., quantum effects in multiple scattering in crystals \cite{Tikhomirov_2019,Tikhomirov_2019_e}).
The spectral distribution of the radiation emitted by
the electron beam is
in good agreement with the experimentally measured spectrum.

Similar sets of simulations have been carried out for the positron
beam of the same energy, transverse size and divergence.
The predictions have been made on the angular distributions
of the deflected positrons as well as on the spectral distribution
of the emitted radiation.
A comparison of the results of simulations for electrons and positrons has
ben carried out.
These results are of interest in connection to the experiments planned to
be carried out at the MAMI facility upon finishing the construction of the
test positron beam (provisionally scheduled for the year 2024).
This activity is carried out within the ongoing project
\cite{TECHNO-CLS}.

%%%%%%%%%%%%%%%%%%%%%%%%%
\subsection*{Acknowledgements}

We acknowledge support by the European Commission through the N-LIGHT
Project within the H2020-MSCA-RISE-2019 call (GA 872196) and the EIC
Pathfinder Project TECHNO-CLS (Project No. 101046458).
PEIA, GRL, MMM and JRS would like to thank NA223LH-INSTEC-003 project from InSTEC-UH. We also acknowledge the Frankfurt Center for Scientific Computing for providing computer facilities.

%%%%%%%%%%%%%%%%%%%%%%%%%%%%%%%
\subsection*{Authors contributions}
All authors were involved in the preparation of the manuscript and contributed equally to this work.
All authors have read and approved the final manuscript.

%%%%%%%%%%%%%%%%%%%%%%%%%%%%%%%
\subsection*{Data Availability Statement}
This manuscript has no
associated data or the data will not be deposited.
[Authors’ comment:
All data generated are included into this published article.
Data will be made available on request.].

%%%%%%%%%%%%%%%%%%%%%%%%%%%%%%%
\subsection*{Conflict of interest}
The authors declare no conflict of interest.

%%%%%%%%%%%%%%%%%%%%%%%%%%%%%%%%%%%%%%%%%%%%%%%%%%%%%%%%
\section*{References}

\bibliography{qmBC}

\end{document}